\documentclass[aps,prb,twocolumn,floatfix,superscriptaddress]{revtex4-1}
\usepackage{times} 
\usepackage{epsfig} 
\usepackage{amsmath}
\usepackage{amssymb} 
\usepackage{graphicx}
\usepackage{dsfont}
\usepackage[colorlinks=true,citecolor=red,linkcolor=blue]{hyperref}

\begin{document}

\title{Spin-selective thermalization plateau in the mass imbalanced Hubbard model}
\author{Liang Du}
\email{LD: liangdu@utexas.edu}
\affiliation{Department of Physics, The University of Texas at Austin, Austin, TX 78712, USA}
\author{Li Huang}
\email{LH: lihuang.dmft@gmail.com}
\affiliation{Science and Technology on Surface Physics and Chemistry Laboratory, P.O. Box 9-35, Jiangyou 621908, China}
\author{Gregory A. Fiete}
\affiliation{Department of Physics, The University of Texas at Austin, Austin, TX 78712, USA}
\begin{abstract}
We use time-dependent non-equilibrium dynamical mean-field theory with weak-coupling auxiliary-field continuous time quantum Monte Carlo as an impurity solver to study the thermalization behavior of the mass-imbalanced single-band Hubbard model after a quench of the Coulomb interaction from the non-interacting limit to a finite positive value. When the Coulomb interaction in our model is increased under equilibrium conditions, the quasi-particle weight for spin-up and spin-down (the mass imbalance) electrons approach zero simultaneously, indicating the absence of a spin-selective Mott transition. By contrast, our out-of-equilibrium study of the mass-imbalanced Hubbard model suggests that there exists the spin-selective dynamical phase transition (one spin orientation undergoes a fast thermalization at its critical Coulomb interaction strength while the other spin orientation shows prethermalization behavior). The spin-selective dynamical phase transition is characterized by the relaxation behavior of the spin-resolved kinetic energy and the spin-resolved momentum-dependent occupation. To make connection with possible experiments, we calculate the spin-resolved two-time optical conductivity, which confirms the spin-selective thermalization plateau. We find the critical Coulomb interaction of each spin orientation for the spin-selective thermalization grows as the mass imbalance decreases.
\end{abstract}
\date{\today}
\maketitle
\section{Introduction}

Research on non-equilibrium, strongly correlated electronic systems has seen dramatic progress on both the theoretical and experimental side in the past decade.\cite{Gull:rmp11, Aoki:rmp14,Fausti:sci11,Conte:sci12,Gandolfi:ps16} In non-equilibrium systems, the observation of hidden quantum states which are not accessible in equilibrium, and the non-equilibrium control of quantum phase transitions in correlated electron systems have attracted much interest.\cite{Tsuji:prl11,Stojchevska:sci14,Mentink:prl14,Mentink:nc15,Eckstein:scirep16,Du:prb17a,Du:prb17b,Puviani:prb16} Two commonly studied scenarios in non-equilibrium correlated electronic systems are the quenched Coulomb interaction in cold atom systems\cite{Manmana:prl07,Rigol:prl07,Kollath:prb07} and the laser driven solid state system.\cite{Mentink:prl14,Mentink:nc15,Eckstein:scirep16,Bukov:prl16} For example, photo-induced transient transitions to superconductivity\cite{Fausti:sci11}, ultrafast switching to a stable hidden quantum state\cite{Stojchevska:sci14} in a layered dichalcogenide crystal of 1T-TaS$_2$, and the dynamical phase transition from anti-ferromagnetic to paramagnetic states\cite{Tsuji:prl13,Tsuji:prb13} illustrate some of the interesting experimentally observed phenomenology.

In addition, the thermalization behavior after a sudden change of one parameter (e.g., the hopping terms after turning on a laser or a Coulomb interaction quench) has attracted much discussion.\cite{Freericks:prl06,Moeckel:prl08,Eckstein:prl09,Amaricci:prb12} Fotso {\it et al.}\cite{Fotso:scirep14} studied the thermalization of the one-band Hubbard model by applying a static electric field.  The authors found there are different thermalization scenarios: (1) Either a monotonic or oscillatory approach to an infinite-temperature steady state; (2) Either a monotonic or oscillatory approach to a non-thermal steady state;  (3) Or evolution to an oscillatory state. By studying the thermalization behavior of an interacting closed system under periodic drive, Abanin {\it et al.}\cite{Abanin:prl15} showed the energy absorption rate decreases exponentially as a function of driving frequency.

By comparison with equilibrium results on the Mott metal-insulator-transition induced by increasing the on-site Coulomb interaction in the single-band Hubbard model, Eckstein {\it et al.}\cite{Eckstein:prl09,Eckstein:prb10} found new behavior out-of-equilibrium--a thermalization plateau after a Coulomb interaction quench in the one-band Hubbard model at half-filling and a dynamical phase transition at a critical Coulomb interaction strength. For small interaction strengths, a prethermalizaton plateau where a quasi-steady state is approached is present on very short time scales, while subsequent thermalization occurs on much longer time scales.  For larger interaction strengths, a collapse-and-revival oscillating behavior of physical observables is observed. Fast thermalization behavior is observed at the critical Coulomb interaction strength. Away from half filling, the dynamical phase transition between these two regimes turns into a crossover.\cite{schiro:prl10}

In this work, we study the thermalization behavior in the mass imbalanced Hubbard model\cite{Dao:pra12,Liu:prb15,Philipp:epjb17,Sekania:prb17} after a Coulomb interaction quench. Here, the mass imbalance implies the hopping amplitude of spin-$\uparrow$ ($V_\uparrow$) and spin-$\downarrow$ ($V_\downarrow$) electrons in the Hubbard model are different. The two limits of mass imbalance are physically motivated and well-studied: In the case $V_\downarrow/V_\uparrow = 0$, the 
spin-$\downarrow$ fermions are fully frozen and we arrive at the Falicov-Kimball model (where we used spin-$\uparrow$ and spin-$\downarrow$ to stand for two the fermion species often called ``$c$" and ``$f$" in the Falicov-Kimball model).\cite{Falicov:prl69,Freericks:rmp03} On the other hand, 
if $V_\downarrow/V_\uparrow = 1$, we recover the mass-balanced Hubbard model. The mass imbalance can be achieved in a cold atom system\cite{Esslinger:arcmp10} by having different atomic species\cite{Taglieber:prl08,Kohstall:nat12} or by generating a spin-dependent hopping through a magnetic field gradient.\cite{Jotzu:prl15} Although the thermalization behavior of Falicov-Kimball and the mass-balanced Hubbard model have been well studied,\cite{Eckstein:prl08,Eckstein:prl09,Eckstein:prb10} it's still unclear what the relaxation behavior will be as the mass imbalance is tuned between the two limiting case above.  This work fills that gap.

In equilibrium, an orbital selective Mott transition has been observed in a  {\em two band} system with different bandwidths\cite{Koga:prl04} and a {\em three band} system with lattice distortion.\cite{Huang:prb12}
Motivated by the orbital selective Mott transition in the {\em multi-band} system, Dao {\it et al.}\cite{Dao:pra12} and Philipp {\it et al.}\cite{Philipp:epjb17} studied the metal-insulator transition in the mass-imbalanced {\em one band} Hubbard model in equilibrium using dynamical mean-field theory combined with different impurity solvers. Their studies suggest that although the spin-up and spin-down electron have different bandwidth, an spin-selective Mott transition can not happen in equilibrium. Further, with increasing mass imbalance, the critical Coulomb interaction for Mott transition decrease monotonically. Based on the equilibrium results for mass imbalanced {\em one band} Hubbard model above, it's natural to ask: (1) Can we find an spin-selective dynamical phase transition in the non-equilibrium system? (2) What is the thermalization behavior dependence on the mass imbalance?
In this work, we show there is indeed a spin-selective dynamical phase transition out-of-equilibrium and we compute the thermalization behavior as a function of the mass imbalance using several observables to illustrate the behavior.

Our paper is organized as follows. In Sec.\ref{sec:model}, we describe the mass imbalanced Hubbard model and illustrate how we calculate several physical observables within dynamical mean-field theory. In Sec.\ref{sec:osdpt}, we characterized the spin-selective dynamical phase transition in the mass imbalanced Hubbard model by presenting the relaxation behavior of the spin-resolved kinetic energy and the momentum distribution for each Fermion species. We also confirmed the spin-selective dynamical phase transition by calculating the spin-resolved optical conductivity. Finally, in Sec.\ref{sec:discussion} we summarize the main conclusions of this work.

\section{Model and Method}
\label{sec:model}
The time-dependent mass-imbalanced single-band Hubbard model at half-filling is given by\cite{Liu:prb15,Philipp:epjb17,Sekania:prb17},
\begin{align}
   H = &\sum_{\langle ij \rangle\sigma} -V_\sigma \left(c^\dagger_{i\sigma} c_{j\sigma}^{} + c^\dagger_{j\sigma} c_{i\sigma}^{}\right)\nonumber\\ 
       &+ U(t) \sum_i \left(\hat n_{i\uparrow}^{} -\frac{1}{2}\right)\left(\hat n_{i\downarrow}^{}-\frac{1}{2}\right),
\label{eq:ham}
\end{align}
where $c^\dagger_{i\sigma}$ ($c_{i\sigma}$) create (annihilate) one electron at site $i$ with spin $\sigma$, $\hat n_{i\sigma} = c^\dagger_{i\sigma} c_{i\sigma}$ is the corresponding number operator, $\langle ij\rangle$ restricts the hopping to nearest neighbors, $V_{\sigma}$ is the corresponding hopping amplitude for a spin $\sigma$ electron ($t$ is reserved for time), and $U(t)$ denotes the time-dependent Coulomb interaction strength between spin-$\uparrow$ and spin-$\downarrow$ fermions on the same site.
Throughout this paper, the system is initially prepared in the ground state of the non-interacting limit ($U_{t<0}=U_i=0$). At $t=0$, the Coulomb interaction is quenched to a constant value $U_{t\geq 0}=U_f>0$ for all later times.
In the following, we set $V_\uparrow = 1$ ($1/V_\uparrow$) as our unit of energy (time) and vary  the mass imbalance $V_\downarrow/V_\uparrow$ between 0 and 1. 

We perform our calculations on the Bethe lattice, which has a semi-elliptic densities of states 
\begin{equation}
\label{eq:dos}
      \rho_\sigma(\epsilon) = \frac{1}{2\pi V_\sigma^2} \sqrt{4V_\sigma^2 - \epsilon^2},
\end{equation}
with half-bandwidth $D_\sigma = 2 V_\sigma$.
The mass-imbalanced Hubbard model (\ref{eq:ham}) can be solved exactly using non-equilibrium dynamical mean field theory (DMFT),\cite{Georges:rmp96,Freericks:prl06,Eckstein:prl09,Gull:rmp11,Aoki:rmp14} which maps the lattice model
self-consistently onto a single-site Anderson impurity model. We use non-equilibrium dynamical mean field theory with continuous time Monte Carlo \cite{Gull:epl08} (CTQMC) as an impurity solver to solve the mass imbalanced Hubbard model at zero temperature.
We enforce a paramagnetic solution and half-filling of both spin-$\uparrow$ and spin-$\downarrow$ electrons. In the mass-balanced Hubbard model, these constraints can be fulfilled by explicitly symmetrizing over the two spin spices and setting the chemical potential to be $\mu = U/2$, respectively. Away from this mass balanced Hubbard model limit, we again enforce half-filling by fixing $\mu = U/2$. However, to ensure the paramagnetic solution at half-filling, we symmetrize the Weiss's functions in the Keldysh time contour using particle-hole symmetry: 
$\mathcal{G}_{0,\sigma}(t, t') = -\mathcal{G}_{0,\sigma}(t',t)$.

\begin{figure}[t]
\includegraphics[width=0.96\linewidth]{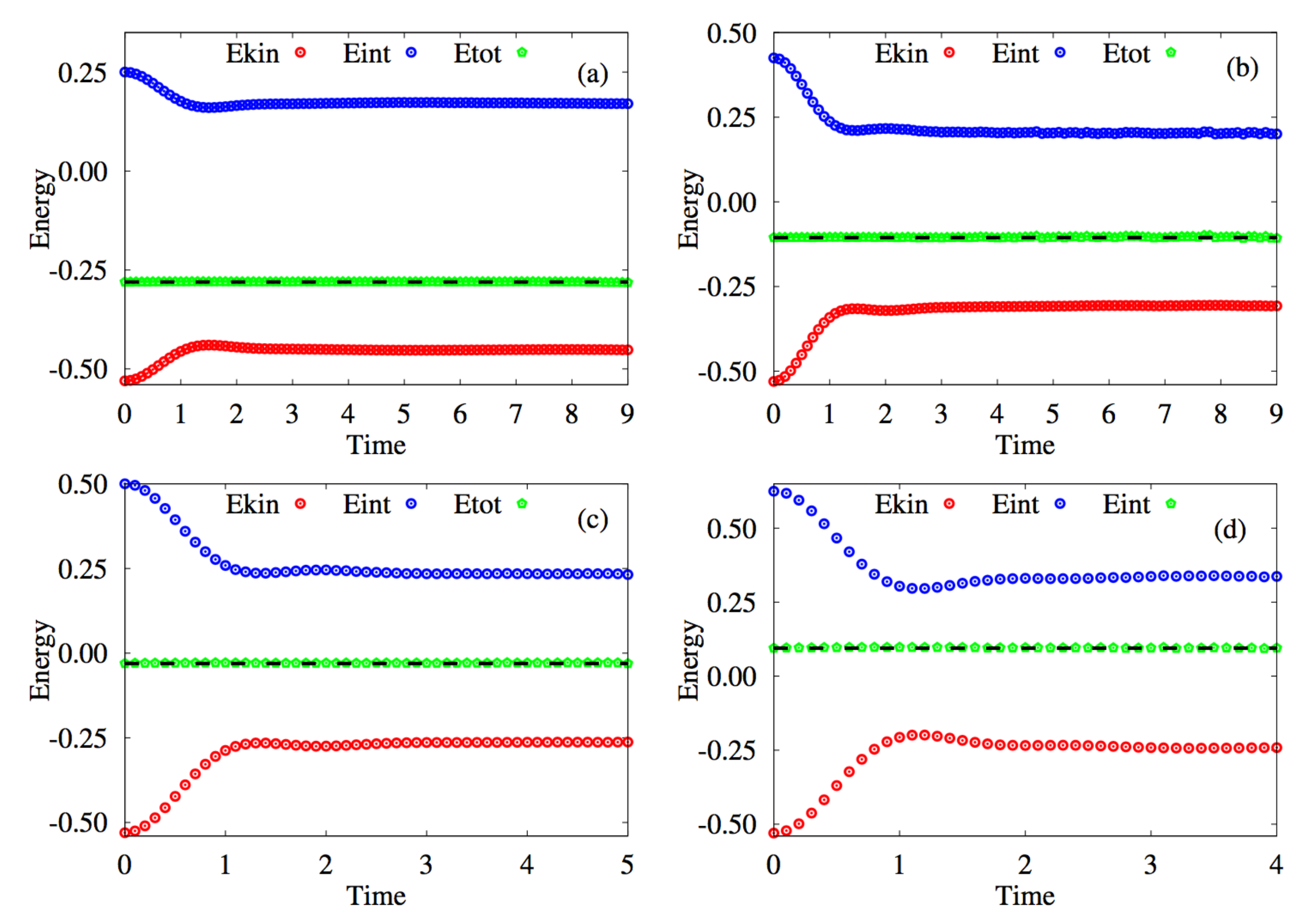}  %
\caption{(Color online) Kinetic (Ekin), Coulomb interaction (Eint), and total energy (Etot) as a function of time for the half-filled mass imbalanced Hubbard model($V_\downarrow/V_\uparrow=1/4$) after an interaction quench from $U_i=0$ to $U_f$ at time $t=0$.
(a) $U_f = 1.0$, (b) $U_f = 1.7$,
(c) $U_f = 2.0$, (d) $U_f = 2.5$. The dashed black line denote the analytical value of total energy using Eq.\eqref{eq:etot}. Since the system is closed, and the Hamiltonian after the quench is time-independent, the total energy is a constant in time.
}
\label{fig:e-vcovf025-small}
\end{figure}

\begin{figure*}[t]
\includegraphics[width=1.00\linewidth]{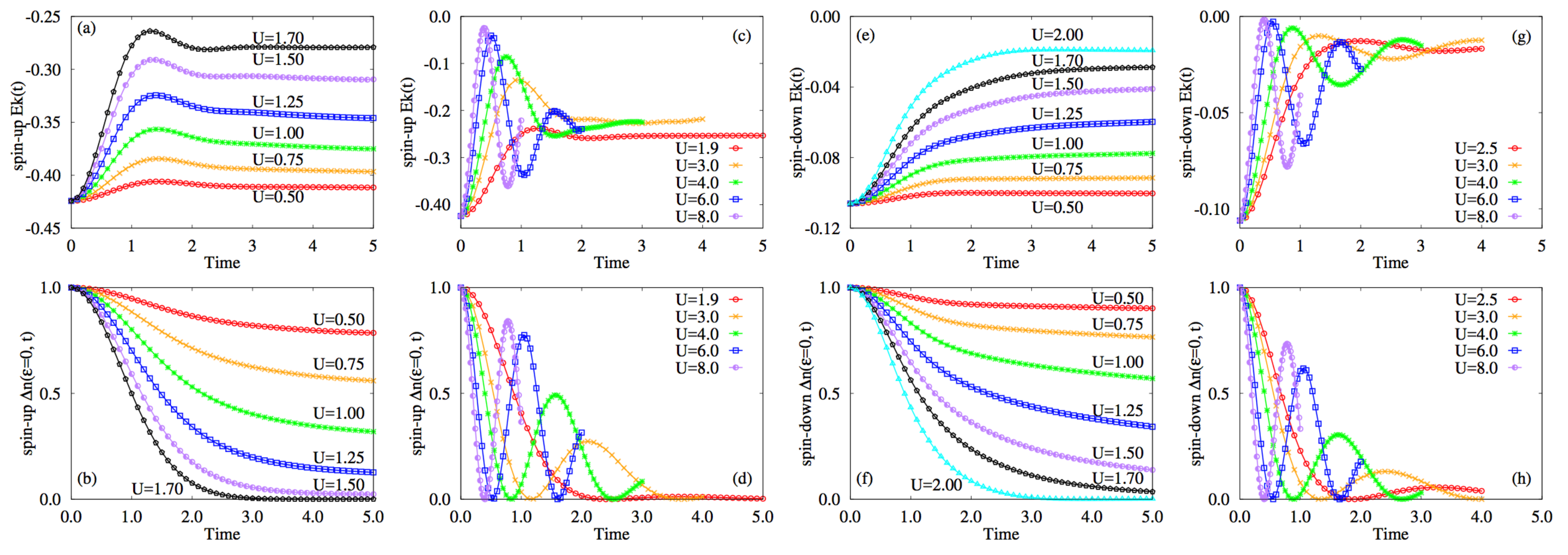}  %
\caption{(Color online) Spin-resolved kinetic energy, and Fermi-surface discontinuity as a function of time for the half-filled mass imbalanced Hubbard model($V_\downarrow/V_\uparrow=1/4$) after an interaction quench from $U_i=0$ to $U_f=U$ at time $t=0$.
(a-b) spin-$\uparrow$ kinetic energy and Fermi surface discontinuity in the relatively weak Coulomb interaction region,
(c-d) spin-$\uparrow$ kinetic energy and Fermi surface discontinuity in the relatively strong Coulomb interaction region, (e-f) spin-$\downarrow$ kinetic energy and Fermi surface discontinuity in the relatively weak Coulomb interaction region, (g-h) spin-$\downarrow$ kinetic energy and Fermi surface discontinuity in the relatively strong Coulomb interaction region. Here the relative weaker or stronger are compared to the critical Coulomb interaction for the spin-selective dynamical phase transition for spin-$\uparrow$ $U_{c\uparrow}=1.7$ and spin-$\downarrow$ $U_{c\downarrow}=2.0$.
}
\label{fig:ssdpt}
\end{figure*}

The expectational value of an observable $\mathcal{O}$ at time $t$ is given by
\begin{align}
      \langle \mathcal{O}(t) \rangle = \frac{1}{Z_0} \mathrm{Tr}[e^{-\beta H(t<0)}U(0,t)\mathcal{O}U(t,0)],
\end{align}
where $Z_0$ is the partition function of the non-interacting Hamiltonian at $t<0$, and $U(t,0) = \mathcal{T}\exp[-i\int_{0}^{t}H(\bar{t})d\bar{t}]$ is the time evolution operator.
The momentum dependent density matrix is written as
\begin{align}
     n_{{\bf k}\sigma}(t) = n_\sigma(\epsilon_{\bf k},t) =  -i G_{{\bf k}\sigma}^{<}(t,t),
\end{align}
where $G_{{\bf k}\sigma}^{<}(t,t)$ is the lesser Green's function at equal time $t$ and momentum independence of the self-energy is assumed.
The time dependent spin-resolved kinetic energy is  given by
\begin{equation}
     E_{\mathrm{kin}}^\sigma(t) = \int d\epsilon_\sigma \rho_{\sigma}^{}(\epsilon_\sigma^{}) \epsilon_\sigma n(\epsilon_\sigma, t).
     \label{eq:ekin}
\end{equation}
The Coulomb interaction energy is given by
\begin{align}
E_{\mathrm{int}}(t) &= U \langle n_{i\uparrow}(t)n_{i\downarrow}(t)\rangle \nonumber\\
     &= 
 -i\int_{\mathcal{C}}d\bar{t} \Sigma_{ii\uparrow}(t,\bar{t}) G_{ii\uparrow}(\bar{t},t) + \langle n_{i\uparrow}(t)\rangle/2,
 \label{eq:eint}
\end{align}
where $\mathcal{C}$ denotes the Keldysh contour.\cite{Eckstein:prb10}
The total energy is 
\begin{equation}
     E_{\mathrm{tot}} = \sum_{\sigma}E_{\mathrm{kin}}^\sigma + E_{\mathrm{int}}.
     \label{eq:etot}
\end{equation}
The Fermi-surface discontinuity is defined as 
\begin{equation}
     \Delta n(t) = n(\epsilon=0^-, t) - n(\epsilon=0^+, t).
\end{equation}
The effective temperature after interaction quench is calculated by numerically solving the equation,\cite{Eckstein:prb10}
\begin{equation}
 E(0^+)=\frac{\mathrm{Tr}\left[H(0^+)e^{-\beta_{\mathrm{eff}}H(0^+)}\right]}
                 {\mathrm{Tr}\left[e^{-\beta_{\mathrm{eff}}H(0^+)}\right]},
                 \label{eq:tempeff}
\end{equation}
where $E(0^+)$ is the same as Eq.\eqref{eq:etot}, and $H(0^+)$ is the Hamiltonian after quench. As the system is fully thermalized, the physical observables will be the same as for the equilibrium system with effective temperature determined by Eq.\eqref{eq:tempeff}, and the same Hamiltonian after quench.
\begin{figure}[t]
\includegraphics[width=0.96\linewidth]{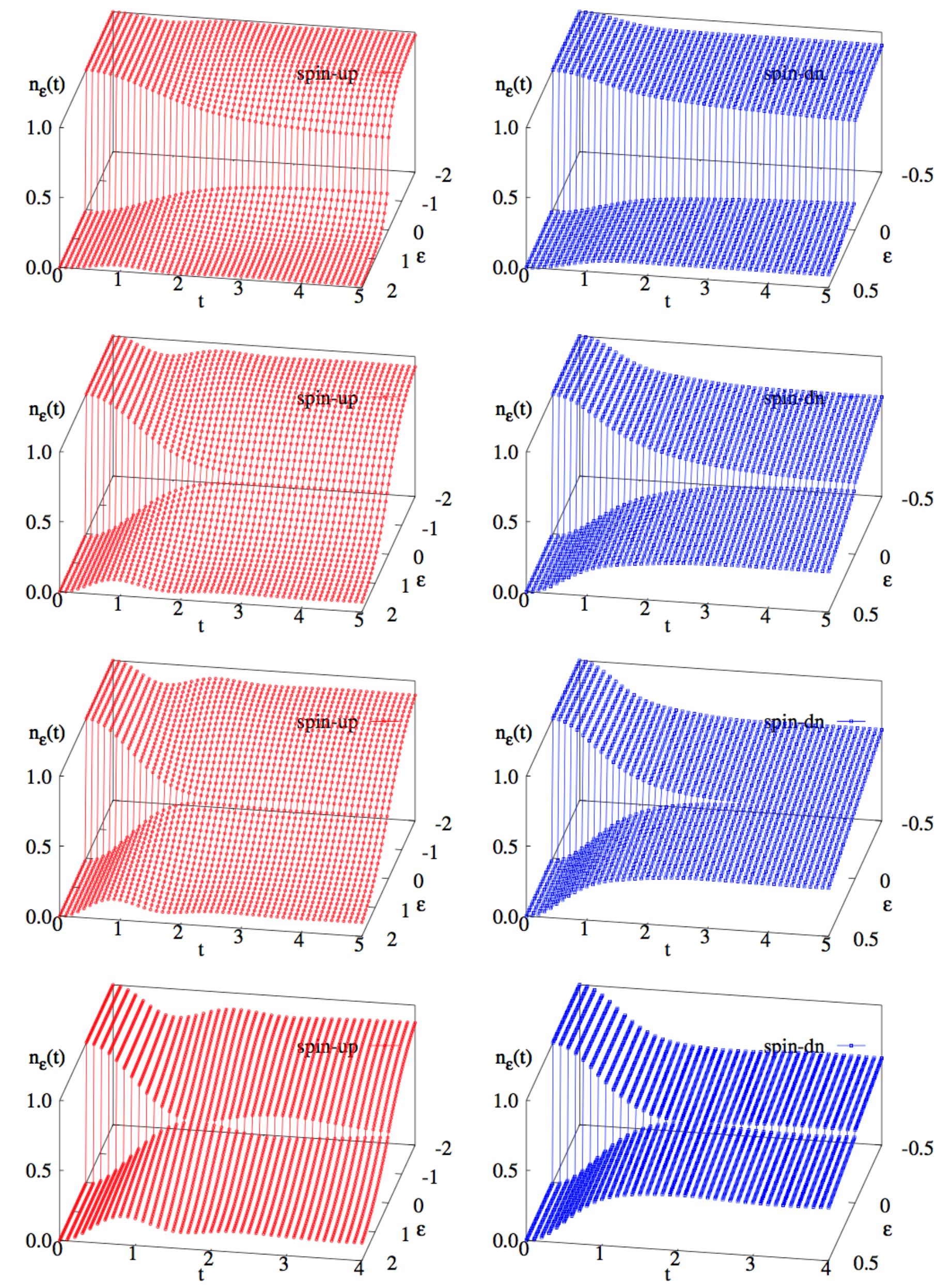}  %
\caption{(Color online) Momentum distribution $n_{\epsilon}(t)$ as a function of time and energy for quenches from $U_i=0$ to $U_f$ for the half-filled mass imbalanced Hubbard model ($V_\downarrow/V_\uparrow=1/4$). 
From top to bottom are the data calculated using non-equilibrium DMFT for $U_f=1.0, 1.7, 2.0, 2.5$, respectively. The left column (red dot) and right column (blue dot) are for spin-$\uparrow$ and spin-$\downarrow$ electrons, respectively.  Note the transition to an oscillatory behavior occurs first for the spin-$\uparrow$ (left column) electrons as the interactions are increased.
}
\label{fig:nk3d-vcovf025-small}
\end{figure}
\begin{figure}[t]
\includegraphics[width=0.96\linewidth]{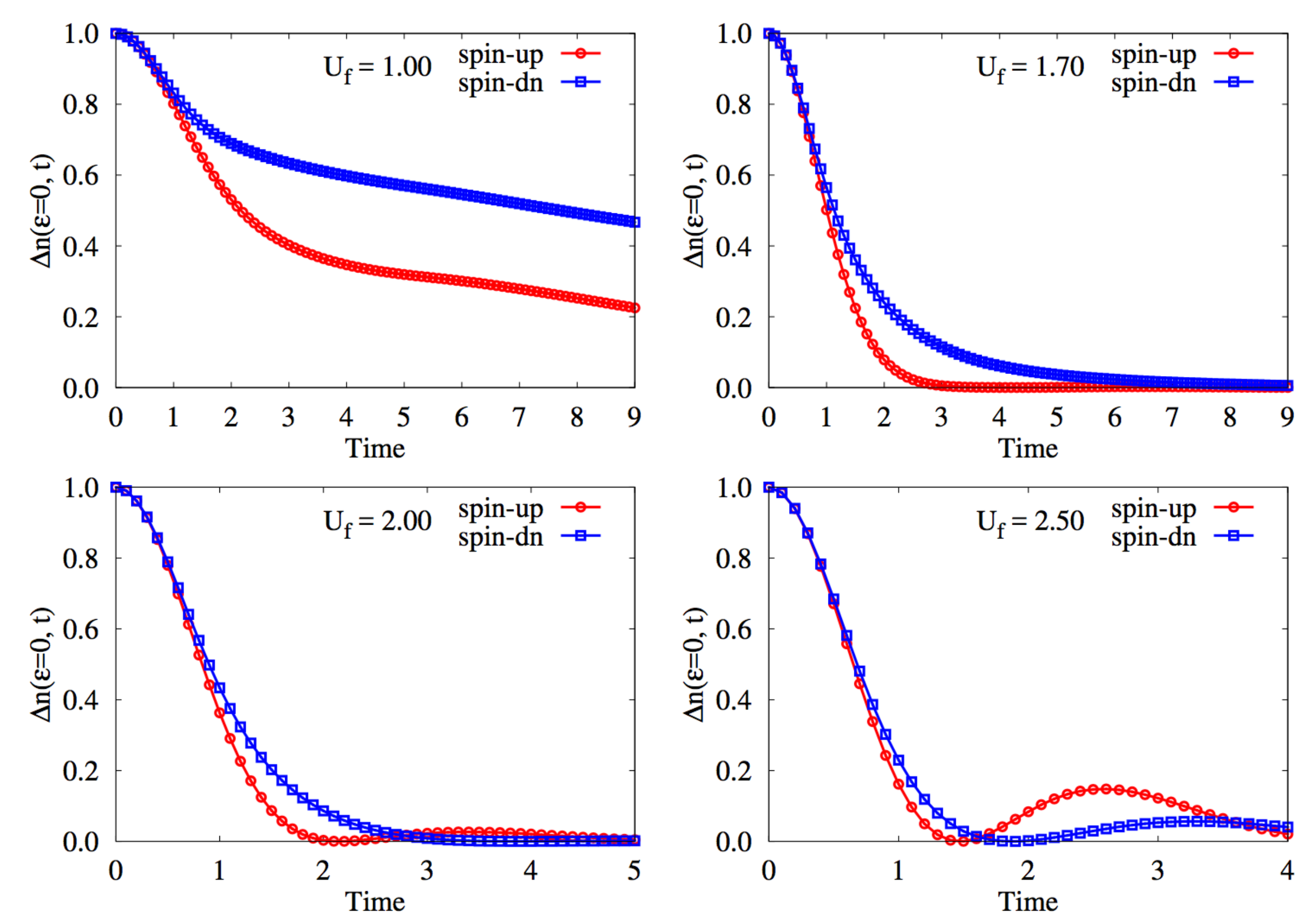}  %
\caption{
(Color online) 
The figure shows the Fermi-surface discontinuity $\Delta n$ (see Fig.\ref{fig:nk3d-vcovf025-small}) as a function of time for different Coulomb interaction $U_f=1.0, 1.7, 2.0, 2.5$.
We fixed $V_\downarrow/V_\uparrow=1/4$ for the half-filed mass imbalanced Hubbard model.
}
\label{fig:nf2d-vcovf025-small}
\end{figure}
\begin{figure}[h]
\includegraphics[width=0.96\linewidth]{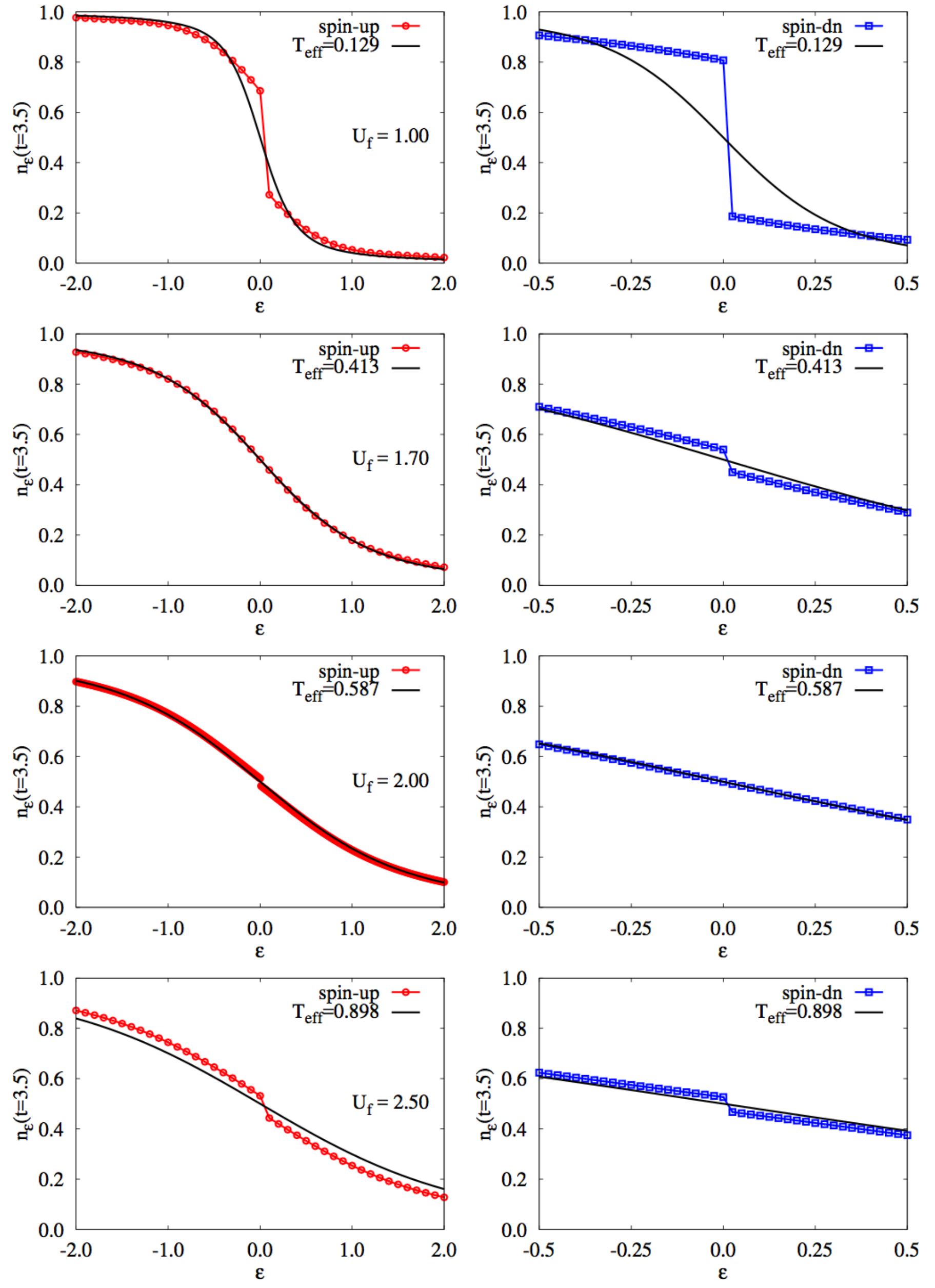}  %
\caption{(Color online) Momentum distribution $n_{\epsilon}(t)$ as a function of energy at fixed time $t=3.5$ for quenches from $U_i=0$ to $U_f$. The dotted points data are calculated using non-equilibrium DMFT.
The solid black line shows the momentum distribution at an effective temperature $T_{\mathrm{eff}}$ calculated using equilibrium DMFT with CTQMC as an impurity solver.
We fixed $V_\downarrow/V_\uparrow=1/4$ for the half-filfed mass imbalanced Hubbard model.
}
\label{fig:nk2d-vcovf025-small}
\end{figure}
\begin{figure}[t]
\includegraphics[width=0.96\linewidth]{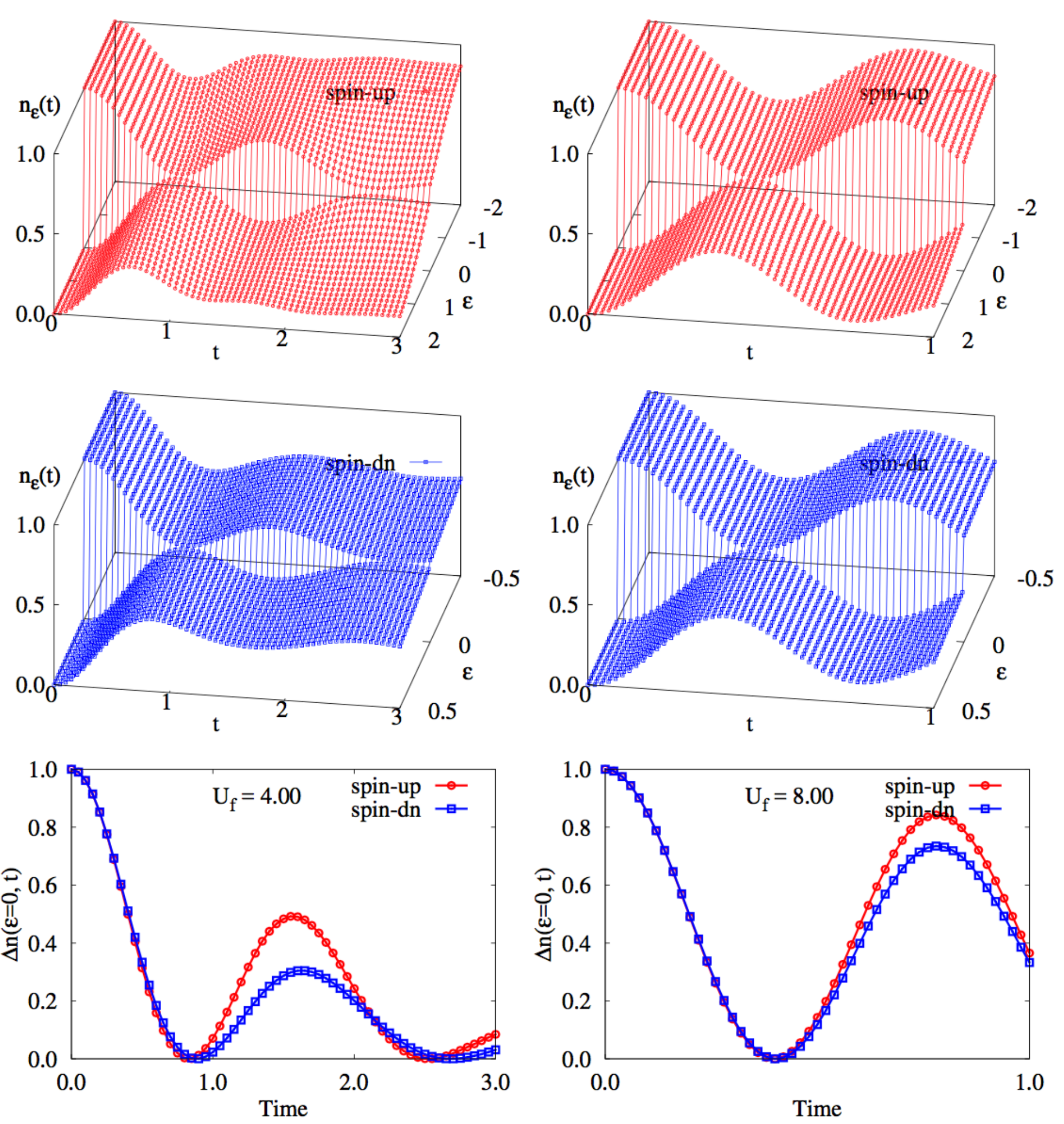}  %
\caption{(Color online) Momentum distribution $n_{\epsilon_k}(t)$ for quenches from $U_i=0$ to $U_f$.
Left column: $U_f=4.0$.
Right column: $U_f = 8.0$. From top to bottom, the data are the momentum distribution as a function of time for spin-$\uparrow$, spin-$\downarrow$ electrons, and the Fermi surface discontinuity, respectively. 
We fixed $V_\downarrow/V_\uparrow=1/4$ for the half-filed mass imbalanced Hubbard model.
}
\label{fig:nk2d-vcovf025-strong}
\end{figure}
\section{Thermalization picture after Coulomb interaction quench}
\label{sec:osdpt}
Since the Hamiltonian after the Coulomb interaction quench $H(t\geq 0)$ is independent of time, the total energy of the closed system will remain constant in time. This character can serve as a check on the reliability of the CTQMC impurity solver applied to the $SU(2)$ symmetry broken case that we consider here. 

In the current case, because the initial Coulomb interaction is zero, $U(t<0)=0$, we have the momentum distribution $n_{\epsilon_{\bf k}}^0=\theta(-\epsilon_{\bf k})$ and double occupancy $\langle n_{\uparrow}n_{\downarrow}\rangle^0 = 1/4$, where $\theta$ is a step function.
The total energy after the Coulomb interaction quench will be 
\begin{align}
    E_{tot}(t=0^+) &= \sum_{\sigma}\int d\epsilon_\sigma \rho_{\sigma}^{}(\epsilon_\sigma^{}) \epsilon_\sigma  n_{\epsilon}^0 + U(t=0^+)/4 \nonumber\\
    &=- 4(V_\uparrow + V_\downarrow)/(3\pi) + U_f / 4,
    \label{eq:etotana}
\end{align}
where we analytically integrate over the Bethe lattice density of states, Eq. \eqref{eq:dos}, up to the Fermi energy.
 
\subsection{Interaction quench for mass imbalance $V_{\downarrow}/V_{\uparrow}=1/4$}
In Fig.\ref{fig:e-vcovf025-small}, we plot the relaxation behavior of the kinetic, Coulomb interaction, and total energy for $U_f=1.0, 1.7, 2.0, 2.5$. The black dashed line is the total energy calculated analytically using Eq.\eqref{eq:etotana}. The dots are the energies calculated using non-equilibrium DMFT with CTQMC as the impurity solver. The analytical and numerical results of the total energy are in good agreement with each other, indicating the CTQMC impurity solver is reliable for the Hamiltonian Eq.\eqref{eq:ham}. The kinetic and Coulomb energy approach the quasi-steady state very fast on a time scale set by $1/U_f$. 
By solving the Eq. \eqref{eq:tempeff} with equilibrium DMFT using CTQMC\cite{Gull:epl08} as the impurity solver, we get the effective temperature for each quenched Coulomb interaction:
\begin{itemize}
     \item $U_f = 1.0; E_{tot} = -0.28052$, $\beta_{\mathrm{eff}} = 7.740$.
     \item $U_f = 1.7; E_{tot} = -0.10552$, $\beta_{\mathrm{eff}} = 2.420$.
     \item $U_f = 2.0; E_{tot} = -0.03052$, $\beta_{\mathrm{eff}} = 1.704$.
     \item $U_f = 2.5; E_{tot} =  +0.09448$, $\beta_{\mathrm{eff}} = 1.114$.
\end{itemize}
If the quenched system arrives at its thermal equilibrium state at sufficiently long time, the temperature of the
state is given by $T_{\mathrm{eff}}=1/\beta_{\mathrm{eff}}$.\cite{Eckstein:prb10} In this paper, we will compare the expectation values of observables after the quench with its thermal equilibrium expectation values at $T=T_{\mathrm{eff}}$ to indicate whether the system is fully thermalized or not.
The effective temperature increases with the final Coulomb interaction because the scattering processes induced by a larger Coulomb interaction will lead to higher a temperature of the system. 
\begin{figure}[t]
\includegraphics[width=0.96\linewidth]{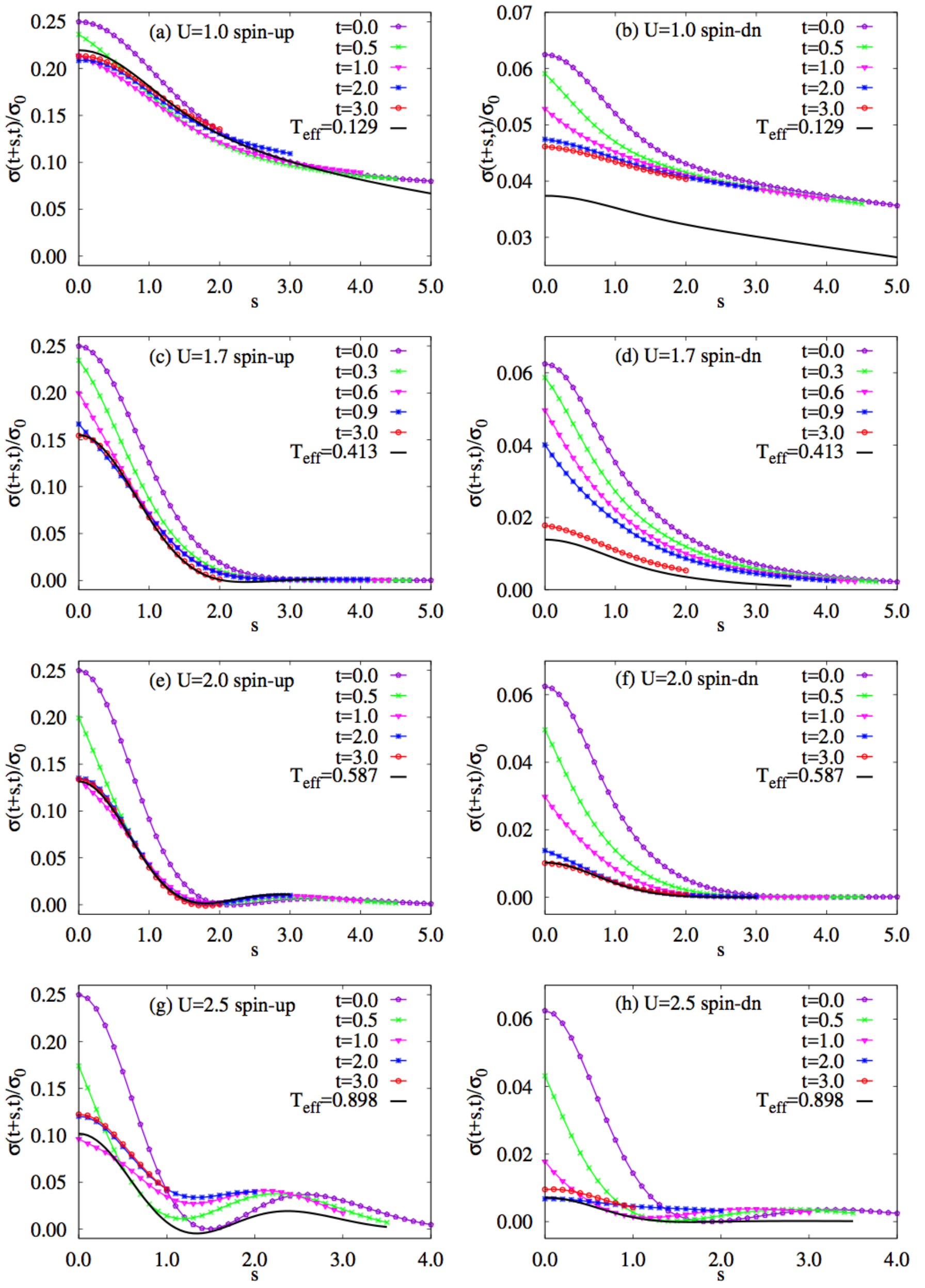}  %
\caption{(Color online) The spin-resolved optical conductivity $\sigma_{\uparrow(\downarrow)}(t+s,t)$ after Coulomb interaction quench to $U=1.0$ (a), $U=1.7$ (b), $U=2.0$ (c), $U=2.5$ (d). The black solid lines shows the optical conductivity in thermal equilibrium at $T_{\mathrm{eff}} = 0.129$ (a-b),$T_{\mathrm{eff}} = 0.413$ (c-d), $T_{\mathrm{eff}} = 0.587$ (e-f), $T_{\mathrm{eff}} = 0.898$ (g-h) .We fixed $V_\downarrow/V_\uparrow=1/4$ for the half-filed mass imbalanced Hubbard model.The left column and right column are data for spin-$\uparrow$ and spin-$\downarrow$ electrons, respectively.
}
\label{fig:opts-vcovf025-small}
\end{figure}
As indicated in Ref.[\onlinecite{Eckstein:prb10}], the Fermi-surface discontinuity of the momentum distribution can be a good criteria to characterize the relaxation after an interaction quench. A finite jump of $\Delta n(\epsilon=0,t)$ clearly indicates the system is not fully thermalized because the thermalized system with finite effective temperature $T_{\mathrm{eff}}$ will have a continuous distribution (that is, no jump) at all the energies. 

Following Ref. [\onlinecite{Eckstein:prl09}], we characterize the spin-selective dynamical phase transition in the quenched mass imbalanced Hubbard model by plotting the evolution behavior of the spin-resolved kinetic energy and Fermi surface discontinuity in Fig.~\ref{fig:ssdpt}. From Fig.~\ref{fig:ssdpt}(a-b), one sees the spin-$\uparrow$ kinetic energy for the relatively weak Coulomb interaction region ($U<U_{c\uparrow}=1.7$) approaches the quasi-stationary state rapidly, reaching it at $t \approx 3$, while the Fermi-surface discontinuity is still evolving over the time scale shown in the figure. For the larger Coulomb interaction region shown in Fig.~\ref{fig:ssdpt}(c-d), both the kinetic energy and the Fermi-surface discontinuity exhibit a decaying collapse-and-revival behavior. However, at the critical Coulomb interaction $U_{c\uparrow}=1.7$, the two quantities undergo a rapid thermalization process and approach the thermalized state already at $t \approx 3$. This sharp crossover between the relatively weak and relatively strong Coulomb interaction behavior is defined as the dynamical phase transition critical point for spin-up electrons. At this critical Coulomb interaction, the spin-$\downarrow$ electron is still evolving up the time scale calculated $t=5$ [shown in Fig.~\ref{fig:ssdpt} (f) with $U=1.7$]. By looking at the kinetic energy and Fermi surface discontinuity for the spin-$\downarrow$ in Fig.~\ref{fig:ssdpt}(e-h), one sees that the critical Coulomb interaction for spin-$\downarrow$ electrons undergo fast thermalization is $U_{c\downarrow}=2.0$. In conclusion, we characterize the dynamical phase transition using the spin-resolved kinetic energy and the Fermi surface discontinuity. The critical Coulomb interaction for different spin orientations is different in the quenched mass imbalanced model. This picture is confirmed by comparison with an equilibrium calculation at an effective temperature and the calculated optical conductivity that will be shown in later sections of this paper.

In Fig.~\ref{fig:nk3d-vcovf025-small} we plot the spin-resolved momentum dependent occupation number as a function of energy and time for different Coulomb interaction $U_f$. 
A cut of Fig.\ref{fig:nk3d-vcovf025-small} to show the Fermi-surface discontinuity is plotted in Fig.~\ref{fig:nf2d-vcovf025-small}. Some general thermalization features are evident in Fig.~\ref{fig:nk3d-vcovf025-small} and Fig.~\ref{fig:nf2d-vcovf025-small}.  At time $t=0$, both spin species are occupied up to the Fermi energy (since the initial condition is a zero-temperature non-interacting system). Hence, the Fermi surface discontinuity is fixed at $\Delta n_{\sigma}=1.0$ for all Coulomb interaction strengths and both the spin species at $t=0$. At very short times, the momentum distributions evolve toward the thermalized state with zero discontinuity at Fermi surface. These features appear to be transient in quenched dynamics. At intermediate time scales, we observed some mixture of the transient and prethermalization pictures. 

At $U_f=1.0$, the Fermi-surface discontinuity decreases monotonically, maintaining a finite value for the time scale shown. This demonstrates the prethermalization picture in which the kinetic and Coulomb interaction energies relax to a quasi-stationary state rapidly (shown in Fig.~\ref{fig:e-vcovf025-small}) while the momentum distribution evolves over a longer time scale. 
By comparing the two spin species, one can see the momentum distribution gap at the Fermi surface is smaller for spin-up particles.
According to the study of Moeckel {\em et al.}\cite{Moeckel:prl08} at weak Coulomb interaction region, $\Delta n_{\sigma}=2Z-1$ for a quasi-stationary state in the weak Coulomb interaction region, where $Z_\sigma$ the quasi-particle weight at zero temperature for spin $\sigma$ calculated in equilibrium. The quasi-particle weight calculated from Ref.[\onlinecite{Dao:pra12,Philipp:epjb17}] shows the particle species with larger bandwidth has smaller quasi-particle weight. 

To further confirm the system is not thermalized on the time scale shown in Figs.~\ref{fig:nk3d-vcovf025-small},\ref{fig:nf2d-vcovf025-small}, we calculate the momentum distribution for the quasi-stationary state with the effective temperature $T_{\mathrm{eff}}=0.129$ obtained by doing equilibrium DMFT with CTQMC as the impurity solver. The comparison of the non-equilibrium results at $t=3$ and the equilibrium data at effective temperature $T_{\mathrm{eff}}$ is shown in Fig.~\ref{fig:nk2d-vcovf025-small} with $U_f=1.0$. The apparent deviation between the data in out-of-equilibrium and in equilibrium ($T_{\mathrm{eff}}=0.129$) indicate the studied system has not reached a thermal state.

\begin{figure}[t]
\includegraphics[width=0.96\linewidth]{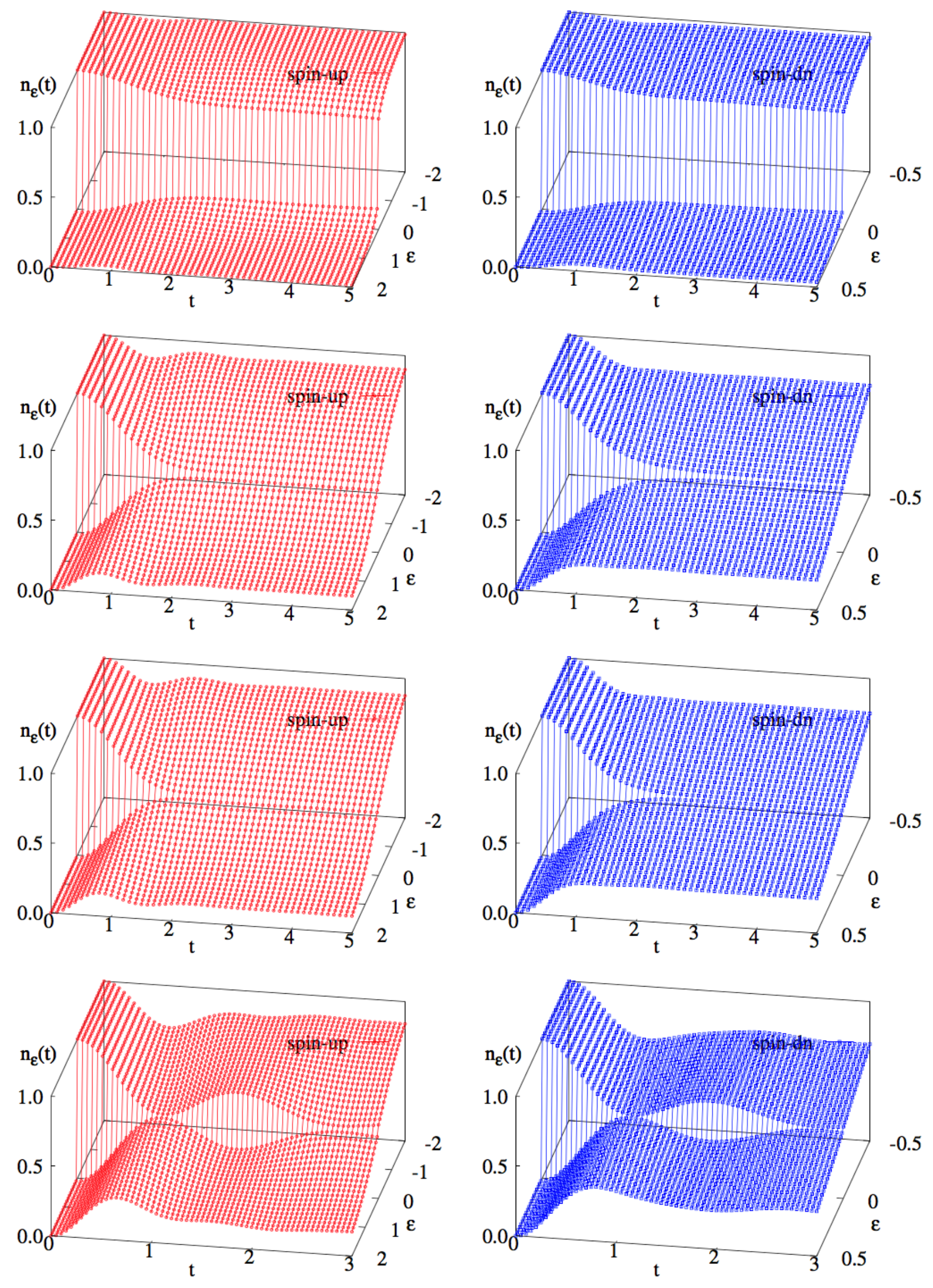}  %
\caption{(Color online) Momentum distribution $n_{\epsilon_k}(t)$ for quenches from $U_i=0$ to $U_f$ for the half-filed mass imbalanced Hubbard model($V_\downarrow/V_\uparrow=1/2$). 
From top to bottom are for $U_f=1.0, 2.2, 2.4, 4.0$, the first and second column show the result for spin-$\uparrow$ and spin-$\downarrow$ electron, respectively.
}
\label{fig:nk3d-vcovf050}
\end{figure}
\begin{figure}[t]
\includegraphics[width=0.96\linewidth]{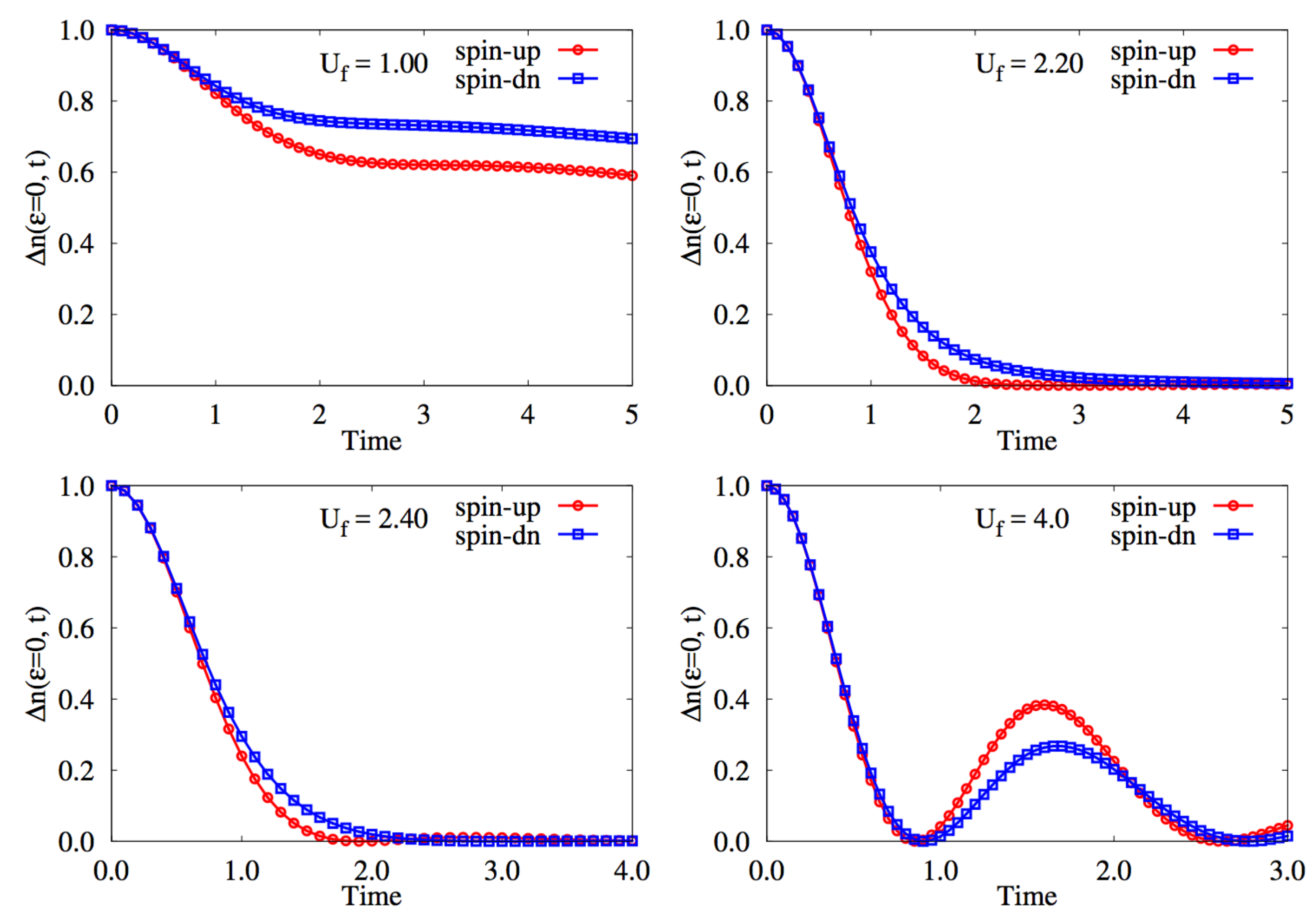}  %
\caption{
(Color online) 
The figure shows the Fermi-surface discontinuity $\Delta n$ as a function of time for different Coulomb interaction $U_f=1.0, 2.2, 2.4, 4.0$ from top to bottom.
We fixed $V_\downarrow/V_\uparrow=1/2$ for the half-filed mass imbalanced Hubbard model.
}
\label{fig:nk2d-vcovf050}
\end{figure}

Further increasing the Coulomb interaction illustrates a very different relaxation behavior. Here we choose three different final Coulomb interactions $U_f=1.7, 2.0, 2.5$. At $U_f=1.7$, the Fermi-surface discontinuity of spin-up particles approaches zero at around $t=2.6$, while the spin-down particles are still gapped up to $t=5.0$. Here the disappearance of the Fermi-surface discontinuity indicates the spin-up particles have approached a thermalized state.  The finite discontinuity of the spin-down particles indicates they are still in a non-thermal state.  This conclusion can be confirmed by doing an equilibrium calculation with the same Hamiltonian after the quench at the effective temperature. The effective temperature $T_{\mathrm{eff}}=0.413$ is calculated through Eq.\eqref{eq:tempeff}.  In Fig.~\ref{fig:nk2d-vcovf025-small} with $U_f=1.7$, for the spin-up particles, the momentum distribution demonstrates a good match for the non-equilibrium and equilibrium results indicating the thermal equilibrium state is reached for this specific spin. However, there exist apparent deviation between the two data for the spin-down particles. This provides clear evidence for the spin selective dynamical phase transition where the spin-$\uparrow(\downarrow)$ is (not) thermalized in some time window.

Continuing to increase the final Coulomb interaction to $U_f=2.0$, we find that the Fermi-surface discontinuity of spin-up particles enters a collapse and revival oscillation with a small amplitude and the spin-down particles approach zero and maintain the same zero value as time increases. Increasing the Coulomb interaction to $U_f=2.7$, we find the two species exhibit clear collapse and revival oscillations, while they have different oscillating periods. From analytical results based on the strong Coulomb interaction limit,\cite{Eckstein:prl09} all observables oscillate with period $2\pi/U_f$. Here we attribute the behavior to be in the intermediate Coulomb interaction regime because the kinetic and Coulomb energy are not oscillating clearly and the oscillations of spin-up and down particles have different periods. To confirm our conclusion, we calculated the momentum distribution in the strong coupling region in Fig.~\ref{fig:nk2d-vcovf025-strong}. Further increasing the final Coulomb interaction up to $U_f=4.0, 8.0$, the momentum distribution for the two spin particles exhibit a collapse and revival oscillation with a period close to each other and finally approach $2\pi/U_f$. 

\subsection{Spin-resolved optical conductivity}
In order to make contact with experiments and to provide more evidence of the spin selective thermalization picture, we compute the two-time spin-resolved optical conductivity $\sigma_{\uparrow(\downarrow)}(t,t')$. In solid state systems, the optical spectroscopy is based on a pump-probe setup in which a strong laser ``pumps" a system out of equilibrium and a weak ``probe" light is used measure the spectroscopy.  Here, for the quenched system, the system is already out of equilibrium, so we need only the weak probe light to determine the linear response of the electrical current in the non-equilibrium state. We use a time-dependent electric field $\delta {\bf E}(\bar{t})$, for the probe light giving a change in the current, 
\begin{equation}
     \delta\langle j(t)\rangle = \int_{-\infty}^t d\bar{t} \sigma(t,\bar{t}) \delta {\bf E}(\bar{t}).
\end{equation}
Details of the calculation of the optical conductivity using non-equilibrium DMFT can be found in Ref.~[\onlinecite{Eckstein:prb08}].  The optical conductivity is measured in units of $\sigma_{0}= 2\rho a^2 e^2 V/\hbar^2$, with $a$ the lattice constant, $\rho$ number of lattice sites per volume and $V$ the volume of the sample. As the system approaches its thermal equilibrium state, the electrical response will become stationary. This can serve as another criteria to check if the system is relaxed to its thermal equilibrium state.\cite{Eckstein:prl08,Eckstein:prb10}

Fig.~\ref{fig:opts-vcovf025-small} shows the spin-resolved two time optical conductivity as a function of the time difference, $s$, with different final Coulomb interaction $U_f = 1.0, 1.7, 2.0, 2.5$. After quenching to $U_f=1.0$ [Fig.\ref{fig:opts-vcovf025-small} (a-b)], the optical conductivity shows a rapid initial relaxation, while the thermal equilibrium state is not approached on the time scale shown in the figure for both spin-$\uparrow$ and spin-$\downarrow$ electrons. The confirms the prethermalization picture supported by the momentum distribution. For the quench to $U_f=1.7$ [Fig.\ref{fig:opts-vcovf025-small} (c-d)], we observe a rapid relaxation of the optical response for the spin-$\uparrow$ electrons. The optical conductivity depends only on the time difference $s$, and is in good agreement with the equilibrium optical conductivity with $T_{\mathrm{eff}} = 0.413$. However, for the spin-$\downarrow$ electrons, 
an apparent deviation of the non-equilibrium data and the equilibrium data suggest the thermal equilibrium state is not reached. The confirms the picture of the spin selective thermalization plateau indicated by the momentum distribution.  Further increasing the final Coulomb interaction to $U_f=2.0$ [Fig.\ref{fig:opts-vcovf025-small} (e-f)], the spin-up electrons show a collapse and revival behavior, while the spin-down electrons exhibit fast relaxation to equilibrium states.
Continuing to increase the final interaction strength [Fig.\ref{fig:opts-vcovf025-small} (g-h)], both spins species exhibit an oscillating character. The deviation from the large interaction limit (with oscillating period $2\pi/U_f$) indicates the Coulomb interaction in still in the intermediate region, which is consistent with the behavior seen in the momentum distribution jump.

\subsection{Dependence of spin selective dynamical phase transition on mass imbalance}
In order to study the dependence of the spin-selective dynamical phase transition, we study the mass imbalanced Hubbard model with a different imbalance ratio,$V_{\downarrow}/V_{\uparrow}=1/2$. The spin-resolved momentum occupation and Fermi surface discontinuity for different Coulomb interactions are shown in Fig.\ref{fig:nk3d-vcovf050} and Fig.\ref{fig:nk2d-vcovf050}, respectively. 
The momentum dependent distribution and Fermi surface discontinuity exhibit very similar behavior compared with $V_{\downarrow}/V_{\uparrow}=1/4$. Prethermalization is illustrated at $U_f=1.0$ with the momentum occupations for the two spin evolving with the time scale shown. At $U_f=2.2$, the spin-$\uparrow$ electron approaches the thermalized state rapidly ($t\approx 2.2$) while the spin-$\downarrow$ electron takes a longer time ($t\approx 4.5$) to thermalize. Further increasing the Coulomb interaction ($U_f=2.4$) will induce an oscillating behavior for the spin-$\uparrow$ electron while the spin-$\downarrow$ electron  thermalizes rapidly.  As the Coulomb interaction moves into the large Coulomb interaction region ($U_f=4.0$), the approximately period $2\pi/U_f$ oscillating behavior is observed.  Here the critical Coulomb interaction for the spin-$\uparrow$ dynamical phase transition is $U_f=2.2$.  The length of the time region where only one-spin species is thermalized is $\Delta t = 4.5-2.2=2.3$ [Fig.\ref{fig:nk2d-vcovf050}]. At the mass imbalance ratio $V_{\downarrow}/V_{\uparrow}=1/4$, the critical Coulomb interaction is $U_f=1.7$. the length of region is $\Delta t = 8.5-2.6=5.9$ [Fig.\ref{fig:nf2d-vcovf025-small}].
Further, as the mass imbalance is reduced, the critical Coulomb interaction for the spin-selective dynamical phase transition is increased.
As one approaches the mass balanced limit, the time widow will disappear\cite{Eckstein:prl09} with critical Coulomb interaction $U=3.2$. Our results are consistent with the equilibrium study of the Mott transition in the mass imbalanced Hubbard model where the critical $U$ increases as the mass imbalance decreases.\cite{Dao:pra12,Philipp:epjb17}

\section{DISCUSSION AND CONCLUSIONS}
\label{sec:discussion}
In this work, we theoretically studied the thermalization behavior in the mass imbalanced Hubbard model after a quench of the Coulomb interaction, which can be realized experimentally in cold atom systems.  We checked the reliability of the impurity solver for the $SU(2)$ symmetry broken case by computing the self-consistency of total energy. We characterized the spin-selective dynamical phase transition by illustrating the relaxation behavior of the spin-resolved kinetic energy and momentum-dependent occupation.  In the weak Coulomb interaction region, a prethermalization picture is observed, where prethermalization is characterized as the kinetic thermalizing to a quasi-stationary state rapidly while the momentum dependent occupation evolves on a longer time scale. In the medium interaction regime, an spin selective thermalization plateau is observed. We find that there exists a critical Coulomb interaction $U_{c\uparrow}$ where the spin-$\uparrow$ particles approach a thermalized state rapidly while the spin-$\downarrow$ particles take longer to evolve. Further increasing the Coulomb interaction in a very small parameter regime to $U_{c\downarrow}$, we find that the spin-$\downarrow$ particles thermalize rapidly to a thermalized state while the momentum dependent occupation of spin-$\uparrow$ particles begin to oscillate in time. Further increasing the Coulomb interaction, the two spin species oscillate, but have different periods. Finally, as one approaches the strong Coulomb interaction limit, the period of the two spin species approach each other and converge to $2\pi/U$. To make contact with experiments, we calculate the spin-resolved two-time optical conductivity. This further supports the spin-selective dynamical phase transition picture. Finally, we study the dependence of spin-selective dynamical phase transition on the mass imbalance, and find that the critical Coulomb interaction increase when the mass imbalance decreases, which is consistent with the equilibrium study on Mott transition in mass imbalanced Hubbard model. 

\section*{Acknowledgements} 
We acknowledge helpful discussions with Qi Chen.  We are grateful to Naoto Tsuji for his generous help while we were coding the non-equilibrium DMFT.  L.H is supported by the Natural Science Foundation of China (No. 11504340). L.D and G.A.F gratefully acknowledge funding from ARO grant W911NF-14-1-0579, NSF DMR-1507621 and NSF MRSEC DMR-1720595.

\bibliography{osdpt.bib}
\end{document}